\documentclass[conference]{IEEEtran}

\usepackage[T1]{fontenc}% optional T1 font encoding
\usepackage[utf8]{inputenc}
\usepackage{cite}
\usepackage{enumitem}

\ifCLASSINFOpdf
  \usepackage[pdftex]{graphicx}
\else
  \usepackage[dvips]{graphicx}
\fi

\usepackage{amsmath,amssymb,amsfonts}
\usepackage{algorithm}
\usepackage{algpseudocode}
\usepackage{url, hyperref}
\usepackage{lastpage}
\usepackage[nolist,nohyperlinks]{acronym}
\usepackage{cite}
\usepackage{xparse}
\usepackage{balance}
\usepackage{xcolor}

\algtext*{EndWhile}% Remove "end while" text
\algtext*{EndIf}% Remove "end if" text
\algtext*{EndFor}% Remove "end for" text

\begin{acronym}
\acro{5G}[5G]{fifth generation}
\acro{3GPP}[3GPP]{3rd Generation Partnership Project}
\acro{NR}[NR]{New Radio}
\acro{mmWave}[mmWave]{millimeter Wave}
\acro{O-RAN}[O-RAN]{Open Radio Access Network}
\acro{RIC}[RIC]{RAN intelligent controller}
\acro{RAN}[RAN]{radio access network}
\acro{RL}[RL]{reinforcement rearning}
\acro{DRL}[DRL]{Deep Reinforcement Learning}
\acro{NP}[NP]{Non-deterministic Polynomial-time}
\acro{SGD}[SGD]{Stochastic Gradient Descent}
\acro{gNB}[gNB]{Next Generation Node Base Station}
\acro{CDF}[CDF]{cumulative distribution function}
\acro{UE}[UE]{user equipment}
\acro{IIoT}[IIoT]{Industrial Internet of Things}
\acro{XR}[XR]{extended reality}
\acro{MEC}[MEC]{Mobile Edge Computing}
\acro{ITU}[ITU]{international telecommunications union}
\acro{SON}[SON]{self-organized network}
\acro{AI}[AI]{artificial intelligence}
\acro{CU}[CU]{central unit}
\acro{DU}[DU]{distributed unit}
\acro{RU}[RU]{radio unit}
\acro{ML}[ML]{machine learning}
\acro{GNN}[GNN]{graph neural networks}
\acro{AWGN}[AWGN]{additive white Gaussian noise}
\acro{RSRP}[RSRP]{received signal reference power}
\acro{DQN}[DQN]{deep Q-network}
\acro{ORAN}[ORAN]{Open Radio Access Network}
\acro{QoS}[QoS]{quality of service}
\end{acronym}

\IEEEoverridecommandlockouts

\begin{document}
\title{\huge Connection Management xAPP for O-RAN RIC: A Graph Neural Network and Reinforcement Learning Approach \vspace{-0.1in}}
\author{}
\author{Oner~Orhan, Vasuki~Narasimha~Swamy,
				Thomas~Tetzlaff,
				Marcel~Nassar,
				Hosein~Nikopour,
				Shilpa~Talwar% <-this % stops a space\\
		\\
		Intel Labs, CA, USA
				\vspace{-0.3in}
\thanks{©2021 IEEE. Personal use of this material is permitted. Permission from IEEE must be obtained for all other uses, in any current or future media, including reprinting/republishing this material for advertising or promotional purposes, creating new collective works, for resale or redistribution to servers or lists, or reuse of any copyrighted component of this work in other works.}}

\maketitle
\thispagestyle{plain}
\pagestyle{plain}
\begin{abstract}
Connection management is an important problem for any wireless network to ensure smooth and well-balanced operation throughout. Traditional methods for connection management (specifically user-cell association) consider sub-optimal and greedy solutions such as connection of each user to a cell with maximum receive power. However, network performance can be improved by leveraging \ac{ML} and \ac{AI} based solutions. The next generation software defined 5G networks defined by the \ac{O-RAN} alliance facilitates the inclusion of \ac{ML}/\ac{AI} based solutions for various network problems. In this paper, we consider intelligent connection management based on the \ac{O-RAN} network architecture to optimize user association and load balancing in the network.
We formulate connection management as a combinatorial graph optimization problem. We propose a deep reinforcement learning (DRL) solution that uses the underlying graph to learn the weights of the \ac{GNN} for optimal user-cell association. 
We consider three candidate objective functions: sum user throughput, cell coverage, and load balancing. 
Our results show up to 10\% gain in throughput, 45-140\% gain cell coverage, 20-45\% gain in load balancing depending on network deployment configurations compared to baseline greedy techniques.
\end{abstract}


\IEEEpeerreviewmaketitle

\acresetall

\section{Introduction}
%\input{introv1}
% Emerging vertical applications such as \ac{IIoT}, \ac{XR}, and autonomous systems impose stringent communication and computation requirements on the infrastructure serving them to deliver seamless, real-time experiences to users \cite{IoT}.
% Cloud computation (as opposed to local, on-device computation) is typically used to support the large computational requirements of these applications. However, the communication latency to the computational cloud server can potentially be very large, resulting in negative user experiences.
% \ac{MEC} addresses this problem by bringing computation resources closer to end users to avoid the typical large delays mentioned above \cite{Abbas}. However, to holistically address the issue, the \ac{RAN} supporting the connection between the user-end devices and the \ac{MEC} server needs to be reliable, high-throughput (data rate) and low latency.
% Hence, the radio network has to be enhanced in \emph{parallel or jointly} with \ac{MEC} solutions to fulfill the new requirements for the emerging applications \cite{Mao}.

Wireless communications systems, both cellular and non-cellular have been evolving for several decades. We are now at the advent of \ac{5G} cellular wireless networks which is considered as the cellular standard to enable emerging vertical applications such as industrial internet of things, extended  reality, and autonomous systems \cite{Andrews}. These systems impose stringent communication and computation requirements on the infrastructure serving them to deliver seamless, real-time experiences to users \cite{IoT}. 
%To address the requirements of the emerging applications, \ac{ITU} initiative and \ac{3GPP} are defining the standardization of the new \ac{3GPP} \ac{NR} that introduces novel designs and technologies to \emph{simultaneously} comply with the requirements for \ac{5G} networks and the novel use cases \cite{3GPP38300}. 
%In addition to a significantly revised core network design and a functional architecture, \ac{5G} \ac{NR} features carrier frequencies up to 52.6~GHz (landing us into the \ac{mmWave} frequency ranges). 
%The large available bands at these \ac{mmWave} frequencies offers the potential of orders of magnitude higher transmission speeds than when operating in the congested bands below 6~GHz and promises 1~Gbps data rate everywhere \cite{Rangan}.
Traditionally, macro base stations provide cellular radio connectivity for devices which has issues such as coverage holes, call drops, jitter, high latency, and video buffering delays. To address these connectivity issues, the \ac{RAN} needs to be brought closer to the end users. This can be achieved through network densification by deploying small cells. The target of this paper is to design and develop a scalable data-driven connection management of dense wireless links \cite{Majdm}.

\begin{figure}[t]
\begin{center}
\centerline{\includegraphics[scale=0.65]{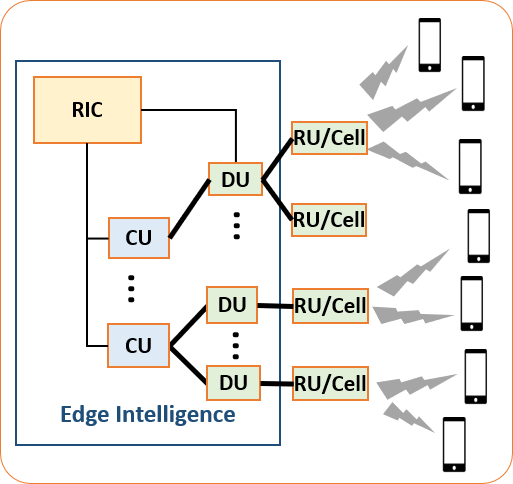}}
\vspace{-0.1in}
\caption{ORAN architecture with distributed controllers located at CU and DU/RU, and intelligence controller RIC} 
\vspace{-0.3in}
\label{fig:top1}
\end{center}
\end{figure}

\subsection{O-RAN architecture}
Typically, front-end and back-end device vendors and carriers collaborate closely to ensure compatibility. The flip-side of such a working model is that it becomes quite difficult to plug-and-play with other devices and this can hamper innovation.
To combat this and to promote openness and inter-operability at every level, \ac{3GPP} introduced RAN dis-aggregation.
In parallel, several key players such as carriers, device manufacturers, academic institutions, etc., interested in the wireless domain have formed the \ac{O-RAN} alliance in 2018 \cite{Operator}. The network architecture proposed by the \ac{O-RAN} alliance is the building block for designing virtualized \ac{RAN} on programmable hardware with radio access control powered by \ac{AI}. 
The main contributions of the \ac{O-RAN} architecture is a) the functionality split of \ac{CU}, \ac{DU} and \ac{RU}, b) standardized interfaces between various units, and c) \ac{RIC}.
The \ac{CU} is the central controller of the network and can serve multiple \ac{DU}s and \ac{RU}s which are connected through fiber links.
A \ac{DU} controls the radio resources, such as time and frequency bands, locally in real time. 
%Every \ac{DU} is equipped with potentially multiple \ac{RU} to establish the radio links. \ac{RU}s at multiple locations provide wireless access between \ac{UE} and network. 
Hence, in the \ac{O-RAN} architecture, the network management is hierarchical with a mix of central and distributed controllers located at \ac{CU} and \ac{DU}s, respectively. 
Another highlight of \ac{O-RAN} architecture is the introduction of a \ac{RIC} that leverages \ac{AI} techniques to embed intelligence in every layer of the \ac{O-RAN} architecture.
More architectural details of ORAN are shown in Figure \ref{fig:top1}.

\subsection{Connection Management}
When a \ac{UE} tries to connect to a network, a network entity has the functionality to provide initial access by connecting the \ac{UE} to a cell. Similarly, when a \ac{UE} moves it needs to keep its connection to the network for smooth operation. These functionalities are called connection management \cite{Tayyab}. 
In addition to managing initial access and mobility, connection management solutions can also be programmed to achieve optimal load distribution. Traditionally, a \ac{UE} triggers a handover request based on wireless channel quality measurements. The handover request is then processed by the \ac{CU}. Connection management in existing solutions is performed using a \ac{UE}-centric approach rather than a context-aware, network-level global approach. One of the common \ac{UE}-centric techniques is \ac{RSRP} based cell-\ac{UE} association. When a \ac{UE} moves away from a serving cell, the \ac{RSRP} from the serving cell will degrade with time while its \ac{RSRP} with a target cell will increase as it gets closer to it. Therefore, a simple \ac{UE}-centric maximum \ac{RSRP} selection approach \cite{Tayyab} can be switching to a new cell when \ac{RSRP} from the target cell is stronger than the current serving cell.

While this greedy approach is simple and effective, it does not take into consideration the network status (local and global). One of the main disadvantage of the greedy approach is the lack of load balancing -- a cell can be heavily loaded\slash congested while other neighboring cells are underutilized, specially with non-uniform user\slash  traffic distribution.
However, \ac{O-RAN} architecture provides the possibility of a more global \ac{RAN} automation by leveraging \ac{ML}-solutions in the \ac{RIC}.
%However, the improved flexibility and agility of the network in the \ac{O-RAN} architecture provides the possibility of a more global \ac{RAN} automation by leveraging \ac{ML}-solutions in the \ac{RIC}.
%One of the most important tasks of the \ac{DU} is load-aware connection management that maximally and equitably utilizes network resources while ensuring smooth operation of the network. 

%\subsection{Related Work}
% While connection management for wireless networks has gained attention, a majority of the work has focused on model-based algorithms. In \cite{Meng}, authors consider cell handover by utilizing \ac{UE} moving speed and direction together with \ac{RSRP} report. A weighted fuzzy self-optimization approach has been proposed in \cite{Alhammadi} for the optimization of cell handover control parameters by considering three levels of fuzzy network features such as low, medium, high traffic load, and slow, moderate, fast \ac{UE} mobility. In \cite{Muhammad}, \ac{UE} association to a cell with lightly loaded data traffic is investigated via cell biasing.  The authors in \cite{Hatipoglu} extend this work to adaptive optimization of connection management parameters defined in \cite{3GPP38300}. The characterization of handover rate is studied in terms of \ac{UE} and cell height in \cite{Arshad}. Overall, these papers study modeling and parameter optimization of connection management for wireless networks.

In \ac{ML}-based optimization framework, dynamic particle swarm optimization is used to improve quality of experience of \ac{UE}s for connection management in \cite{Fang}. In \cite{Alrabeiah}, a visual-data-assisted handover optimization is considered by using neural networks. A more proactive approach by predicting obstacles to associate \ac{UE}s to new cells before link disconnection is proposed in \cite{Wang}. %In \cite{Sun}, multi-arm bandit approach is considered to minimize the number of handover events in the network. 
In a more distributed learning framework, authors in \cite{Sana} investigate \ac{UE} throughput maximization using multi-agent reinforcement learning which considers independent handover decisions based on local channel measurements. Similarly, \cite{Khosravi} studies the same problem using deep deterministic reinforcement learning algorithm to solve the resulting non-convex optimization problem. The above machine learning algorithms do not utilize  structure of wireless networks for the design of neural network architecture, and hence, may  have  performance loss from wireless network dynamics.

%The above model based algorithms suffer from modeling error and sub-optimal performance. The existing machine learning algorithms do not utilize structure of wireless networks for the design of neural network architecture, and hence, may have performance loss from variable wireless network size and dynamics. 
In this paper, we consider an \ac{AI}-based framework for load-aware connection management which incorporates structure of wireless networks into the neural network architecture. Specifically, we focus on the problem of handover management using \ac{GNN} and \ac{RL} as our main tools. To achieve intelligent and proactive connection management, we abstract the \ac{O-RAN} network as a graph, in which cells and \ac{UE}s are represented by nodes and the quality of the wireless links are given by the edge weights. To capture the load-awareness, edge and node labels reflecting features, such as instantaneous load conditions, channel quality, average \ac{UE} rates, etc. are considered and the proposed joint \ac{GNN}-\ac{RL} framework is applied to enable intelligent user handover decisions.

\section{System Model}
\label{sys}
\begin{figure}[ht]
\begin{center}
\centerline{\includegraphics[scale=0.65]{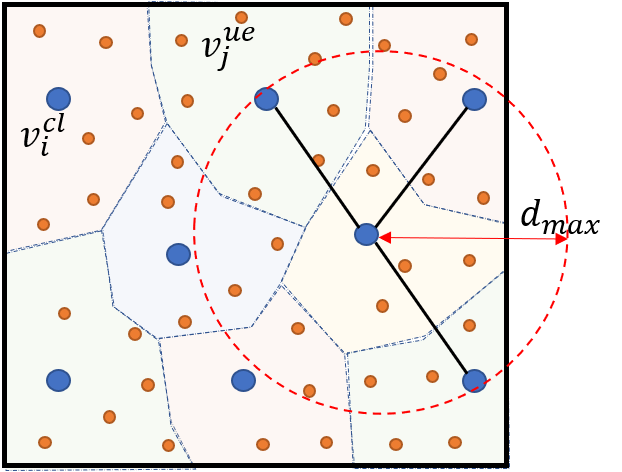}}
\vspace{-0.1in}
\caption{An example network abstraction as a graph: blue circles are cells and orange circles are \ac{UE} nodes.}
\label{fig:top2}
\end{center}
\vspace{-0.2in}
\end{figure}

%Broadly, \ac{O-RAN} networks consist of \ac{DU}s, \ac{UE}s and a single central \ac{RIC}. The \ac{DU}s have wired backhaul links to the central \ac{O-RAN}-\ac{RIC} and supports wireless access to the \ac{UE}s. 
In this paper, we consider an \ac{O-RAN} network consisting of $N$ cells (we assume that every \ac{RU} represents a cell) and $M$ \ac{UE}s as a graph $\mathcal{G}=(\mathcal{V},\mathcal{E})$. The set of cell nodes are $\mathcal{V}^{cl} = \{v_0^{cl},...,v_{N-1}^{cl}\}$ and the set of \ac{UE} nodes are $\mathcal{V}^{ue} = \{v_0^{ue},...,v_{M-1}^{ue}\}$ with the set of all nodes in the network given by $\mathcal{V}= \mathcal{V}^{cl} \cup \mathcal{V}^{ue}$. The edges $\mathcal{E}^{ue} = \{e_{v_i^{cl},v_j^{ue}} | v_i^{cl} \in \mathcal{V}^{cl},  v_j^{ue} \in \mathcal{V}^{ue}\}$  of $\mathcal{G}$ are wireless links between \ac{UE}s and cells. Although  all cells are directly connected to a \ac{RIC} in a tree structure, we consider virtual edges between cells (\ac{RU}) to convey information about their \ac{UE} connectivity and local graph structure. The virtual edges $\mathcal{E}^{cl}=\{e_{v_i^{cl},v_j^{cl}} | v_i^{cl},v_j^{cl} \in \mathcal{V}^{cl}\}$ between two cells can be defined according to the Euclidean distance such that there is a link between two cells if the Euclidean distance between them is smaller than $d_{max}$ (this is just one way to define the edges). We denote the set of \ac{UE} nodes connected to a specific cell, $v_i^{cl}$, as $\mathcal{C}(v_i^{cl})=\{v_j^{ue} | e_{v_i^{cl},v_j^{ue}} \in \mathcal{E}^{ue} ,\forall j \}$. An example \ac{O-RAN} network abstraction is given in Figure \ref{fig:top2}. As shown in the figure, the cell-\ac{UE} connections are depicted as shaded clustering around cells, and cell-cell virtual connection graph is decided according to the Euclidean distance. 

The links between \ac{UE}s and cells are dynamic and they depend on the mobility of the \ac{UE}s. In communication theory, the channel capacity quantifies the highest information rate that can be sent reliably (with a small probability of error). A rough estimate of the single-input single-output channel capacity between base station and user device with \ac{AWGN} at the receiver is given by
\begin{align}\label{eq:ch_cap}
c(v_i^{cl},v_j^{ue}) &= \log_2\left(1 + \frac{P(v_i^{cl},v_j^{ue})}{N_0}\right), \text{ bits/sec}
\end{align}
where $N_0$ is the noise power and $P(v_i^{cl},v_j^{ue})$ is \ac{RSRP} at $v_j^{ue}$ from cell $v_i^{cl}$.  The above estimate is more accurate if we assume that the interference from neighboring cells is negligible and links are beamformed (especially for mmWave). We also disregard the interference from non-assigned % oner: what is non-assigned node
nodes since mmWave frequency narrow beams are known to be power-limited rather than being interference-limited. We assume that each \ac{UE} measures \ac{RSRP}s from close-by cells and reports them to the \ac{RIC}. Then, the \ac{RIC} decides on the connectivity graph between cells and \ac{UE}s according to a desired network performance measure. 
%The connectivity graph can be considered as \ac{UE} clustering with \ac{DU} indexes. 
We consider the following performance measures at the network:
\begin{itemize}
\item \textbf{Sum throughput:} Given a graph $\mathcal{G}$, the network throughput is defined as a total data rate it can deliver to the \ac{UE}s in the network. The throughput is computed as follows:
\begin{align}\label{eq:metric1}
U_{th} (\mathcal{G}) =\sum_{i=0}^{N-1} \sum_{j=0}^{M-1} \frac{c(v_i^{cl},v_j^{ue})}{|\mathcal{C}(v_i^{cl})|}, \text{ bits/sec}
\end{align}
Here, we consider equal resource allocation between \ac{UE}s connected to the same cell. 
\item \textbf{Coverage:} \ac{UE}s can be classified as cell-centric or cell-edge depending on the data rate they get. A user is considered as cell-edge if its rate is below a chosen threshold. In general, this threshold value is chosen to be the $5^{\text{th}}$ percentile of the all \ac{UE} rates in the network and is the \emph{coverage} of the network. Higher cell-edge user rate improves network coverage and reduces coverage holes. 
\begin{align}\label{eq:metric2}
U_{cov} (\mathcal{G})= \inf \left\{y : F(y)<0.05 \right\},
\end{align}
where $y \in \left\{ \frac{c(v_i^{cl},v_j^{ue})}{|\mathcal{C}(v_i^{cl})|}, \forall i, j\right\}$, and $F(\cdot)$ is \ac{CDF}. 
\item \textbf{Load balancing:} In communication networks, various fairness metric are considered to ensure equitable allocation of resources \cite{Shi}. In this work, we consider \textit{Jain's index} to quantitatively measure fair resource allocation between users. The  \textit{Jain's index} is defined as,
\begin{align}\label{eq:metric3}
U_{Jain} (\mathcal{G})= \frac{ \left[\sum_{i=0}^{N-1} |\mathcal{C}(v_i^{cl})| \right]^2}{ M\sum_{i=0}^{N-1} |\mathcal{C}(v_i^{cl})|^2 }
\end{align}
\end{itemize}

In our optimization problem, we aim to find the optimal graph $\mathcal{G}^*$ leading to the best \ac{UE} and cell association such that a combination of the above performance measures is maximized. The optimal network topology/graph $\mathcal{G}^*$ is given by:
\begin{equation}\label{eq:Objective}
\mathcal{G}^* = \arg\max_{\mathcal{G}}U(\mathcal{G}).
\end{equation}
where $U(\mathcal{G})$ can be a weighted combination of performance measures defined above.

\section{Graph Neural Networks}\label{sec:GNN}
Graph Neural Networks are a framework to capture the dependence of nodes in graphs via message passing between the nodes. Unlike deep neural networks, a \ac{GNN} directly operates on a graph to represent information from its neighborhood with arbitrary hops. This makes \ac{GNN} an apt tool to use for wireless networks which have complex features that cannot be captured in a closed form. In this paper, we consider \ac{GNN}-based approach by incorporating cell-\ac{UE} relationship between nodes as well as channel capacities over the edges. 

For a given network with a set of $N$ cells and $M$ \ac{UE}s, we define two adjacency matrices: $\mathbf{A}_{cl} \in \{0,1\}^{N\times N}$ for the graph between cells and $\mathbf{A}_{ue} \in \{0,1\}^{N\times M}$ for the graph between \ac{UE}s and cells, as follows:
\begin{align}
\mathbf{A}_{cl}(i,j) &= \begin{cases}
											1 & \text{if } e_{v_i^{cl},v_j^{cl}}  \in \mathcal{E}^{cl}\\
											0 & \text{o.w.}
										 \end{cases} 
\end{align}
\vspace{-0.15in}
\begin{align}
\mathbf{A}_{ue}(i,j) &= \begin{cases}
											1 & \text{if } e_{v_i^{cl},v_j^{ue}} \in \mathcal{E}^{ue}\\
											0 & \text{o.w.}
										 \end{cases} 
\end{align}
We consider a $L$-layer GNN that computes on the graph.
We define the initial nodal features of the cells and \ac{UE}s as $(\mathbf{X}_{cl,1}^{(0)}, \mathbf{X}_{cl,2}^{(0)})$ and $\mathbf{X}_{ue}^{(0)}$, respectively. The initial nodal features are functions of reported channel capacities and data rates at the cell and \ac{UE}. We define $\mathbf{C}\in \mathbb{R}^{N\times M}$ as channel capacity matrix with elements $c(v_i^{cl},v_j^{ue})$, and $\mathbf{R}\in \mathbb{R}^{N\times M}$ as user rate matrix with elements $\frac{c(v_i^{cl},v_j^{ue})}{|\mathcal{C}(v_i^{cl})|}$ for a given cell-\ac{UE} connectivity graph. We calculate input features as follows:
\begin{align}\label{eq:InFeatureDU}
\mathbf{X}_{cl,1}^{(0)} &= [\mathbf{A}_{cl}\mathbf{R}\mathbf{1}_M || \mathbf{R}\mathbf{1}_M ] \in \mathbb{R}^{N\times 2}\\
\mathbf{X}_{cl,2}^{(0)} &= [\mathbf{A}_{ue} \mathbf{R}^T\mathbf{1}_N ||
\mathbf{C}\mathbf{1}_M]  \in \mathbb{R}^{N\times 2}\\
\mathbf{X}_{ue}^{(0)} &= [\mathbf{C}^T\mathbf{1}_N || \mathbf{R}^T\mathbf{1}_N  ] \in \mathbb{R}^{M\times 2}
\end{align}
where $[\cdot||\cdot]$ is vector concatenation operator and $\mathbf{1}_M$ and $\mathbf{1}_N$ are all-ones vector of size $M$ and $N$, respectively. All the above latent features capture either node sum rate or sum rates of neighboring cells or channel capacity/data rate in the case of \ac{UE}s. These are selected as the features since they capture the information relevant to making connectivity decisions.

At every layer, the \ac{GNN} computes a $d$-dimensional latent feature vector for each node $v_i^{cl},v_j^{ue} \in \mathcal{V}$ in the graph $\mathcal{G}$. The latent feature calculation at layer $l$ can be written as follows:
\begin{align}\label{eq:GNN}
\mathbf{H}_{cl}^{(l)} & = \sigma \left(\mathbf{X}_{cl,1}^{(l)} \mathbf{W}_{1}^{(l)}\right) +\sigma \left(\mathbf{X}_{cl,2}^{(l)} \mathbf{W}_{2}^{(l)}\right)\in \mathbb{R}^{N\times d}\\
\mathbf{H}_{ue}^{(l)} &= \sigma \left(\mathbf{X}_{ue}^{(l)} \mathbf{W}_{3}^{(l)}\right)\in \mathbb{R}^{M\times d}\\
\mathbf{X}_{cl,1}^{(l+1)} &= \mathbf{A}_{cl}\mathbf{H}_{cl}^{(l)} \in \mathbb{R}^{N\times d}\label{eq:GNN_eq3}\\
\mathbf{X}_{ue}^{(l+1)} &= \mathbf{A}_{ue}^T\mathbf{H}_{cl}^{(l)} \in \mathbb{R}^{M\times d}\label{eq:GNN_eq4}\\
\mathbf{X}_{cl,2}^{(l+1)} &= \mathbf{A}_{ue} \mathbf{H}_{ue}^{(l)} \in \mathbb{R}^{N\times d}\label{eq:GNN_eq5}
\end{align}
In the above equations, $\mathbf{W}_{k}^{(0)} \in \mathbb{R}^{2\times d}$ and $\mathbf{W}_{k}^{(l)} \in \mathbb{R}^{d\times d}$ (for $l > 0$), $k=1,2,3$, are neural network weights, $l$ is the layer index of \ac{GNN}, and $\sigma(\cdot)$ is a non-linear activation function. Note that $\mathbf{H}_{cl}^{(l)}$ and $\mathbf{H}_{ue}^{(l)}$ are auxiliary matrices which represent sum of hidden features of cell-cell and cell-\ac{UE} connectivity graphs. Equations (\ref{eq:GNN_eq3})-(\ref{eq:GNN_eq5}) represent a spatial diffusion convolution neural network \cite{gnnsurvey}. 
The $L$-layer \ac{GNN} essentially repeats the above calculation for $l=0,1,..,L-1$. Through this, features of the nodes are propagated to other nodes and will get aggregated at distant nodes. This way each node's feature will contain information about its $L$-hop neighbors, as the embedding is carried out $L$-times. 

We combine the feature vectors at the last layer of \ac{GNN} to get a scalar-valued score for $\mathcal{G}$. We sum the output layer of \ac{GNN} over cells, $\mathbf{H}_{cl}^{(L-1)}$, which makes the score calculation invariant to permutation over nodes, before passing it to single layer fully connected neural network. We get network score of the graph $\mathcal{G}$ as follows:
\begin{align}\label{eq:score}
Q(\mathcal{G}) &= \sigma \left(\mathbf{1}_N^T \mathbf{H}_{cl}^{(L-1)} \mathbf{W}_{4}\right) \mathbf{w}_{5},
\end{align}
where $\mathbf{1}_N^T$ is the all-ones vector of size $N$, $\mathbf{W}_{4}\in \mathbb{R}^{d\times d}$ is the fully connected neural network weight matrix, and $\mathbf{w}_{5}\in \mathbb{R}^{d\times 1}$ is the vector to combine neural network output, linearly. 

Once the \ac{GNN} computations are complete, the score of $\mathcal{G}$, $Q(\mathcal{G})$, will be used to select the best connection graph among subset of feasible graphs. The procedure to learn the optimal weights $\mathbf{W}_{k}^{(l)}$, $\forall k,l$, and $\mathbf{w}_{5}$ is described in the next section.

\section{Deep Q-learning Algorithm}\label{sec:DQNAlgorithm}
We propose a deep Q-learning approach \cite{rlref}, in which a $Q$-function is learned from cell and \ac{UE} deployment instances and the corresponding reward we get from the network environment. The advantage of the proposed \ac{GNN} formulation as the neural network for the $Q$-function is that \ac{GNN} is scalable to different graph sizes and can capture local network features with variable numbers of cells and \ac{UE}s.
To make the best selection for \ac{UE} connectivity, we need to learn the right $Q$-function. As the $Q$-function is captured through the \ac{GNN}, this translates to learning the parameters of the \ac{GNN} which we do through sequential addition of new cell-\ac{UE} connections to partially connected graph.

The state, action, and reward in the deep \ac{RL} framework are defined as follows:
\begin{itemize}[leftmargin=*]
\item \textbf{State} $s_t$: The state is defined as the current graph $\mathcal{G}_t$ containing the cells and connected \ac{UE}s at iteration $t$ as well as input features of nodes $\mathbf{X}_{cl}^{(0)}$ and $\mathbf{X}_{ue}^{(0)}$. The start state can be considered as partially connected network with connected and unconnected \ac{UE}s. The terminal state $s_T$ is achieved when all the \ac{UE}s in the network are connected.
\item \textbf{Action} $a_t$: The action $a_t = \mathcal{G}_t \cup e_{v_i^{cl},v_j^{ue}}$ at step $t$ is to connect an unconnected \ac{UE} to one of the cells.
\item \textbf{Reward} $r(s_t,a_t)$: The reward at state $s_t$ after selecting action $a_t$ is 
\begin{equation}\label{eq:Reward}
r(s_t,a_t) = U(\mathcal{G}_t) - U(\mathcal{G}_{t-1}),
\end{equation} i.e., the reward is defined as the change in the network utility function after connecting a new \ac{UE}. In section \ref{training}, we provide various reward functions for the performance measures given in Section \ref{sys}.
\item \textbf{Policy} $\pi(a_t|s_t)$: We use a deterministic greedy policy, i.e., $\pi(a_t|s_t)= \arg\max_{a_t} Q(s_t,a_t)$ with $\epsilon$-greedy exploration during training. Here, $Q(s_t,a_t)$ is defined in Eq.~(\ref{eq:score}) with $\mathcal{G}_{t} = s_{t} \cup e_{v_i^{cl},v_j^{ue}}$ 
\end{itemize}

\begin{algorithm} % enter the algorithm environment
\caption{DQN Based Connection Management} % give the algorithm a caption
\label{alg:DRL} % and a label for \ref{} commands later in the document
\begin{algorithmic}[1] % enter the algorithmic environment		
		\State Randomly initialize weights $\mathbf{W}_{k}^{(l)}$, $\forall k,l$, and $\mathbf{w}_{5}$ (normally distributed with random small variance)
		\State Define number of network deployment scenario $I$, episodes $K$, and \ac{GNN} layers $L$
		\For{deployment $i=1:I$}
				\State For deployment $i$, get $\mathcal{G}_0$, $\mathbf{X}_{cl, 1}^{(0)}$, $\mathbf{X}_{cl, 2}^{(0)}$ and $\mathbf{X}_{ue}^{(0)}$
				\State Initialize state $s_0 = \mathcal{G}_0$ (cell-cell connectivity graph
				\State defined in Section \ref{sys} and some initial cell-\ac{UE} pairings)
				\For{step $ t =0:T-1$}
						\State Take action $a_t \sim \pi(a_t|s_t)$
						\State Move to the next state $\mathcal{G}_{t}\leftarrow s_t \cup e_{v_i^{cl},v_j^{ue}}$
						\State Compute input features $\mathbf{X}_{cl,1}^{(0)}$, $\mathbf{X}_{cl,2}^{(0)}$ and $\mathbf{X}_{ue}^{(0)}$
						\State Observe reward $r(s_t,a_t)$
						\State Perform $L$-layer \ac{GNN} in (\ref{eq:score}) to compute 
						\State $Q(s_t,a_t)$ and the following:
						\State \vspace{-0.3cm}\begin{equation}\label{eq:QUpdate}y = \gamma \max_{a_{t+1}}Q(s_{t+1}, a_{t+1}) + r(s_t,a_t)\end{equation}\vspace{-0.3cm}
						\State where $\gamma$ is discount factor
						\State Update parameters:
						\State $\mathbf{W}_{k}^{(l)}\leftarrow \mathbf{W}_{k}^{(l)}+\alpha\left(y - Q(s_t,a_t)\right)\nabla_{\mathbf{W}_{k}^{(l)}}Q(s_t,a_t)$
						\State $\mathbf{w}_{5}\leftarrow \mathbf{w}_{5}+\alpha\left(y - Q(s_t,a_t)\right)\nabla_{\mathbf{w}_{5}}Q(s_t,a_t)$
						\State where $\alpha$ is learning rate
						\State Use $\epsilon-$greedy policy:
            \[ \pi(a_{t+1}|s_{t+1}) = \begin{cases}
											\text{random cell-\ac{UE} pairing, i.e., } & \\ 
											\hspace{0.7in} a_{t+1} = e_{v_i^{cl},v_j^{ue}},& \text{w.p. } \epsilon\\
											\arg\max_{a_{t+1}} Q(s_{t+1},a_{t+1}),& \text{o.w.}
										 \end{cases}
						\]
			   \EndFor
			\EndFor
\end{algorithmic}
\vspace{-0.1in}
\end{algorithm}

Algorithm \ref{alg:DRL} describes the proposed \ac{DQN} approach. First, the parameters are initialized and defined for each deployment. In each step $t$, one \ac{UE} $a_t = e_{v_i^{cl},v_j^{ue}}$ is connected by following the $\epsilon$-greedy policy $\pi(a_t|s_t)$, with $\epsilon$ being the exploration rate. Here, the number of steps $T$ is given by the termination state $s_T$. The graph $\mathcal{G}_t$ is updated, so that the next step $s_{t+1}$ is obtained. The new nodal input features $\mathbf{X}_{cl}^{(0)}$ and $\mathbf{X}_{ue}^{(0)}$ are calculated every time the graph is updated, and the reward $r(s_t, a_t)$ is calculated for each selected action. The $L$-layer \ac{GNN} computation provides the score for each state and action pair. 

Then, to learn the neural network weights $\mathbf{W}_{k}^{(l)}$, $\forall k,l$, and $\mathbf{w}_{5}$, $Q$-learning updates parameters by performing \ac{SGD} to minimize the squared loss
$E\{\left(y - Q(s_t,a_t)\right)^2\}$,
with $y$ being defined in Eq.~(\ref{eq:QUpdate}) and $\gamma$ being the discount factor. Algorithm \ref{alg:DRL} reflects the training phase. Once the training is completed, the neural network weights are not updated and they are used directly to obtain the actions for unseen deployment instances.

\section{Implementation and Evaluation}\label{sec:results}
\subsection{xApp Implementation at O-RAN RIC}\label{xapp}
\ac{O-RAN} defines xApp as an application designed to run on the near real-time operation at the \ac{RIC} \cite{Operator}. These xApps consist of several microservices which get input data through interfaces between \ac{O-RAN} \ac{RIC} and \ac{RAN} functionality, and provides additional functionality as output to \ac{RAN}. This section mainly addresses the methods to make the xApp scalable and provides an overview on how the connection management algorithm is deployed and realized in \ac{O-RAN} architecture. Even though it also works for initial access, we consider the proposed \ac{GNN}-\ac{RL} based connection management algorithm for handover application in which mobile users in the network request for new cell connections. We refer to the request for a new cell connection as a \emph{handover event}. A UE continuously measures the RSRPs from its surrounding cells. If certain conditions are met (as defined in the 3GPP standards), the UE reports the measured RSRPs for a handover request. When the \ac{O-RAN} \ac{RIC} receives a handover event, the \ac{GNN}-\ac{RL} algorithm makes new connection decisions to balance the load of the network. 

We expect that the \ac{O-RAN} \ac{RIC} consists of 100s of cells and 1000s of \ac{UE}s. The large scale \ac{O-RAN} deployment will result in a large network topology graph $\mathcal{G}$ and which increases the processing latency and complexity of the \ac{GNN}-\ac{RL} inference. We consider two solutions to reduce dimension of \ac{GNN}-\ac{RL} inference. First, we consider a local sub-graph of the \ac{O-RAN} network around a handover requested \ac{UE}. This local sub-graph includes only those cells whose \ac{RSRP} is reported by \ac{UE} that has issued the handover request and the $L-$hop  neighbors of the these cells in the virtual cell-cell connection graph as defined in Section \ref{sys}. Here, $L$ is the number of layer of \ac{GNN} as defined in Section \ref{sec:GNN}. Second, we classify each \ac{UE} in the network as either a cell-edge or a cell-center \ac{UE}. The cell-edge \ac{UE}s are defined as the \ac{UE}s that are close to the boundary of the cell's coverage as shown in Figure \ref{fig:top2}. We mark the \ac{UE} as a cell edge \ac{UE} if the difference between the strongest and the second strongest \ac{RSRP} measurements is less than a given threshold e.g. $3$dB. The remaining \ac{UE}s are marked as cell-center \ac{UE}s since their strongest \ac{RSRP} measurement is larger than their other measurements, and hence, does not need a new cell connection. Therefore, the initial connectivity graph $\mathcal{G}_{0}$ of the \ac{GNN}-\ac{RL} includes an edge between a cell and a \ac{UE} if it is a cell-center \ac{UE}. We refer to the set of cell-edge \ac{UE}s in the sub-graph as \emph{reshuffled \ac{UE}s}. The solution proposed above enables us to reduce the total action space of the \ac{RL} algorithm by reducing the number of reshuffled \ac{UE}s, $T$, in the initial connectivity graph $\mathcal{G}_0$ in Algorithm \ref{alg:DRL}.

\subsection{Training}\label{training}
To showcase the benefits of the proposed \ac{GNN}-\ac{RL} algorithm in various use cases and applications, we train the \ac{GNN} with two different reward functions described below. Then, we evaluate the performance with metrics given in Section \ref{sys}.

For data intensive applications where maximizing throughput is more important, we consider the sum throughput utility function given in Eq.~ (\ref{eq:metric1}) to calculate reward as follows:
\begin{align}\label{eq:Reward1}
r(s_t,a_t) &= U_{th}(\mathcal{G}_t) - U_{th}(\mathcal{G}_{t-1}).
\end{align}

For applications that prioritize fairness among users, we consider the following reward function which is weighted sum of improvement in total network throughput and the smallest user rate at each cell in the network (captured by the second term in the equation below): 
\begin{align}\label{eq:Reward2}
r(s_t,a_t) &= U_{th}(\mathcal{G}_t) - U_{th}(\mathcal{G}_{t-1}) \nonumber \\
&+ \frac{\lambda}{|\mathcal{V}^{cl}|} \sum_{v_i^{cl} \in \mathcal{V}^{cl}}  \min_{v_j^{ue}:e_{v_i^{cl},v_j^{ue}} \in \mathcal{E}^{ue}} c(v_i^{cl},v_j^{ue})
\end{align}
Note that the last term in the above equation tries to maximize the minimum user rate. Increasing the minimum user rate helps to maximize the network coverage given in Eq. (\ref{eq:metric2}) and fairness given in Eq. (\ref{eq:metric3}) by closing the rate gap between users.

We consider uniformly distributed cells and \ac{UE}s in a hexagonal network area. The consideration of random deployment is useful to generalize inference performance to many real world cases such as varying city block sizes, rural or urban areas, hot spots at stadiums and concerts. We follow 3GPP network settings and channel models \cite{3GPP38300}. The cell transmit power is 33 dBm. The carrier frequency of channel is 30GHz with the large scale channel parameters and 100MHz channel bandwidth \cite{Rangan}. In the network, each \ac{UE} measures the \ac{RSRP} from its current serving cell and its three closest cells, and reports the measurements back to the \ac{O-RAN} \ac{RIC}.

For training the \ac{GNN}, we collect $1000$ deployment scenarios with $6$ cells and $50$ \ac{UE}s. We set the diameter of hexagonal area to $500m$ and select 6 cells in the area which corresponds to about 37 cells per $km^2$. For the \ac{GNN} architecture, we have $L=2$ layers, and $d=8$ dimensions per layer. For the reinforcement learning algorithm, we consider exploration rate $\epsilon = 0.1$, learning rate $\alpha=0.1$ and discount factor $\gamma=1$. Additionally, we consider experience buffer of size $8$ to reduce the impact of correlation between consecutive \ac{UE} association.

\subsection{Numerical Results}
We compare \ac{GNN}-\ac{RL} solution with the maximum \ac{RSRP} benchmark algorithm. In the benchmark algorithm, each \ac{UE} is associated with a cell from which it receives the strongest \ac{RSRP}. As discussed in Section \ref{sys}, the benchmark algorithm is \ac{UE}-centric and greedy. To show the scalability and robustness benefits of the \ac{GNN}-\ac{RL} approach, we collect $50$ different deployment scenarios for different number of cells and \ac{UE}s and network densities.

\begin{figure}[ht]
\vspace{-0.1in}
\begin{center}
\centerline{\includegraphics[scale=0.55]{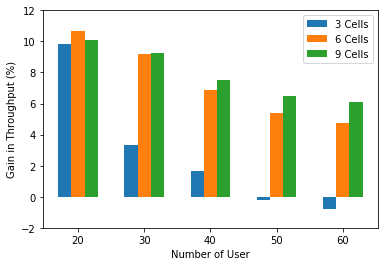}}
\vspace{-0.1in}
\caption{Throughput gain of \ac{GNN}-\ac{RL} with various network sizes}
\label{fig:eval1}
\end{center}
\vspace{-0.3in}
\end{figure}

In Fig. \ref{fig:eval1}, we depict the relative gain of throughput defined in (\ref{eq:metric1}) of \ac{GNN}-\ac{RL} approach over the maximum \ac{RSRP} algorithm. In this case, the \ac{GNN} weights are obtained using reward function given in Eq.~(\ref{eq:Reward1}). As shown in the figure, we observe up to $10 \%$ gain when the number of \ac{UE}s is small and as the number of users increases the gain drops. This is expected because when the number of users is small, each user gets larger share from the network, and a connection decision made by the \ac{GNN}-\ac{RL} approach has more impact on the performance. On the other hand, as the network size scales up with the number of cells while keeping diameter of hexagonal network area the same, we also observe more gain in performance which shows scalability and robustness benefits of the \ac{GNN} architecture.

\begin{figure}[ht]
\begin{center}
\vspace{-0.1in}
\centerline{\includegraphics[scale=0.55]{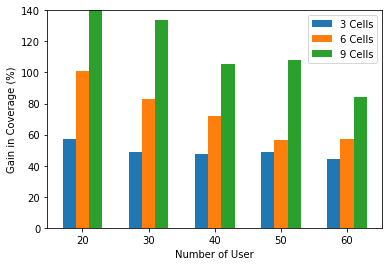}}
\vspace{-0.1in}
\caption{Coverage gain of \ac{GNN}-\ac{RL} with various network sizes}
\label{fig:eval2}
\end{center}
\vspace{-0.3in}
\end{figure}

\begin{figure}[ht]
\begin{center}
\vspace{-0.1in}
\centerline{\includegraphics[scale=0.55]{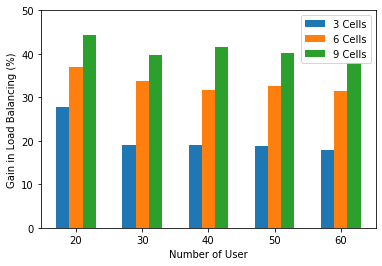}}
\vspace{-0.1in}
\caption{Load balancing gain of \ac{GNN}-\ac{RL} with various network sizes}
\label{fig:eval3}
\end{center}
\vspace{-0.3in}
\end{figure}

In Fig.~\ref{fig:eval2} and \ref{fig:eval3}, we show the relative gain of coverage and load balancing defined in Eq~(\ref{eq:metric2}) and (\ref{eq:metric3}), respectively, of \ac{GNN}-\ac{RL} approach over the maximum \ac{RSRP} algorithm. Here, we train the \ac{GNN} with the reward function given in Eq.~(\ref{eq:Reward2}). We observe similar trends as in Fig.~\ref{fig:eval1}. However, the relative gains in coverage and load balancing is much larger than the throughput gain which shows the importance of \ac{GNN} based solution for handover applications.

\begin{figure}[ht]
\begin{center}
\vspace{-0.1in}
\centerline{\includegraphics[scale=0.55]{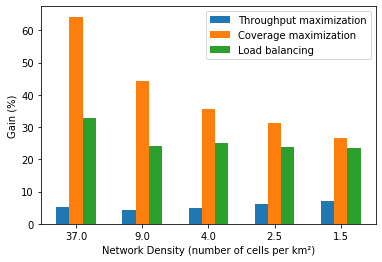}}
\vspace{-0.1in}
\caption{Gain of \ac{GNN}-\ac{RL} with various network densities}
\label{fig:eval4}
\end{center}
\vspace{-0.3in}
\end{figure}

Fig.~\ref{fig:eval4} shows the benefit of \ac{GNN}-\ac{RL} approach to varying network densities in terms of number of cell per $km^2$ while keeping the average number of \ac{UE}s per cell the same. As argued before, we train the neural network only for the scenario with $37$ cells per $km^2$ network density and use trained model to test different network densities. We observe more gain in coverage as network gets denser because when network is dense, cell edge users have multiple good cell selection options and \ac{GNN}-\ac{RL} approach makes better decisions compared to greedy cell selection. Additionally, high performance gains in different network densities show that the \ac{GNN}-\ac{RL} approach is robust to any network deployment scenario.

% Finally, we also compare the performance of the \ac{GNN}-\ac{RL} approach on independent network data generated by Open Networking Foundation in \cite{onf}. The RAN simulator in \cite{onf} has a network density of 1.5 cells per $km^2$ with mobile \ac{UE}s following random trajectories over Google Maps \cite{Google}. The RAN simulator in \cite{onf} reports time correlated \ac{RSRP} measurements according to \ac{UE} mobility in the real world. We observe about $25.6\%$ coverage gain in RAN simulator in \cite{onf}. Similarly, our simulator provides $26.47\%$ coverage gain as shown in Fig.~\ref{fig:eval4} which shows that the proposed \ac{GNN}-\ac{RL} approach is robust to different network deployment models.

\section{Conclusion}\label{sec:conc}
In this paper, we introduce connection management for \ac{O-RAN} \ac{RIC} architecture based on \ac{GNN} and deep \ac{RL}. The proposed approach considers the graph structure of the \ac{O-RAN} architecture as the building block of neural network architecture and use \ac{RL} to learn the parameters of the algorithm. The main advantage of the algorithm is that it can consider local network features to make better decisions to balance network traffic load while network throughput is also maximized. We also demonstrate that the proposed approach is scalable and robust against different network scenarios, and outperforms the existing \ac{RSRP} based algorithm.

\end{document}